\documentclass[12pt]{article}
\begin{document}
\title{Incompleteness of trajectory-based interpretations of quantum mechanics \footnote{\copyright IOP Publishing 2004}}
\author{Michael J. W. Hall\\
Theoretical Physics, IAS, \\ Australian National
University,\\
Canberra ACT 0200, Australia}
\date{}
\maketitle



\begin{abstract}
Trajectory-based approaches to quantum mechanics include the 
de~Broglie-Bohm interpretation and Nelson's stochastic interpretation.  It is shown that the usual route to establishing the validity of such interpretations, via a decomposition of the Schr\"{o}dinger equation into a continuity equation and a modified Hamilton-Jacobi equation, fails for some quantum states.  A very simple example is provided by a quantum particle in a box, described by a wavefunction initially uniform over the interior of the box. For this example there is no corresponding continuity or modified Hamilton-Jacobi equation, and the spacetime dependence of the wavefunction has a known fractal structure.  Examples with finite average energies are also constructed.
\end{abstract}


\section{Introduction}

The formalism of standard quantum mechanics is very different from that of classical mechanics, as are the generic phenomena described by each theory. This gives rise to a number of well known interpretational issues when one tries to integrate classical and quantum aspects of the world.  The Copenhagen interpretation of quantum mechanics remains foremost for most physicists in resolving such issues, and adequately explains the empirical content of the standard formalism.  However, a number of alternative interpretations exist, and can be valuable in providing (i) reasonably coherent pictures for thinking about fundamental quantum phenomena such as interference and entanglement; (ii) means for marrying the microscopic with the macroscopic (in contrast to the enforced separation specified by the Copenhagen interpretation);  and (iii) a variety of starting points for extending or modifying the standard quantum formalism.  

One class of alternative interpretations is distinguished by retaining the classical concept of spacetime trajectories for quantum particles.  Thus, for example, in the de~Broglie-Bohm interpretation particles follow trajectories in configuration space determined by a guiding wave, providing an underlying deterministic (but nonlocal) picture of quantum evolution \cite{bohm, holland}.  A second example is Nelson's stochastic interpretation, in which particles follow non-differentiable trajectories in configuration space determined by a stochastic generalisation of Newton's second law \cite{nelsonpaper,nelsonbook}. 

An important claim made by proponents of such trajectory-based interpretations is that they reproduce {\it all} predictions of standard quantum mechanics, at least for the case of nonrelativistic particles moving under velocity-independent potentials \cite{bohm, holland,nelsonpaper,nelsonbook}.  This general reproducibility is a necessary feature of any {\it complete} interpretation of the quantum formalism.  Indeed, the dBB and stochastic interpretations appear capable of going beyond the standard formalism, as they have even been applied to non-normalisable wavefunctions lying outside the Hilbert space of possible quantum states (eg, sections 4.10, 4.11 of Ref.~\cite{holland}, and page 88 of 
Ref.~\cite{nelsonbook}).  

However, the aim of this note is to show that such trajectory-based interpretations in fact do {\it not} apply to all quantum states, and hence are formally incomplete.  Moreover, since some of the states in question have finite average energies, it is suggested that such interpretations may also be {\it physically} incomplete.  Similar difficulties arise for any interpretation based on continuity and modified Hamilton-Jacobi equations, including the hydrodynamic and exact uncertainty approaches to quantum mechanics 
\cite{holland,hydro,exact}.

The results are based on a subtle property of the Schr\"{o}dinger equation in the position representation, discussed in section 2.  In particular, for certain states of quantum particles, this equation cannot be partitioned into separate terms involving spatial and temporal derivatives of the wavefunction respectively.  A simple example is provided by a wavefunction initially uniform over the interior of some region and vanishing elsewhere (eg, a plane wave incident on a slit, or a particle confined to a box with maximal position entropy).  

The existence of such states formally arises as a consequence of the unboundedness of the corresponding Hamiltonian operator - this operator cannot in fact be directly applied to a large class of wavefunctions, even though, paradoxically, these wavefunctions and their evolution are themselves perfectly well defined.  Further, as shown in section 3, such wavefunctions can have finite average energies, and hence can, in principle, be physically prepared from finite resources.

An important consequence of unboundedness is that the Schr\"{o}dinger equation cannot always be decomposed into a continuity equation and a modified Hamilton-Jacobi equation.  As discussed in section 4, the nonexistence of such a decomposition for certain states leads to an incompleteness of trajectory-based interpretations for these states.

The lack of continuity and modified Hamilton-Jacobi equations, for the particular case of a particle in a box described by an initially uniform wavefunction, is connected with the known fractal structure of this wavefunction in almost all spatial and temporal directions \cite{berry}.  In section 5 it is conjectured that, more generally, the nonexistence of these equations corresponds to either the wavefunction or its spatial derivative having a fractal structure (where the latter case corresponds to examples having finite average energies).
 
Conclusions are given in section 6.

\section{A subtlety of the Schr\"{o}dinger equation}

Attention will primarily be restricted to quantum systems comprising a single nonrelativistic spin-zero particle. The corresponding Hilbert space is then given by the set of square-integrable complex functions on the configuration space of the particle.  It is typically assumed, in what follows, that the configuration space is one-dimensional (results can easily be generalised to higher dimensions).

Consider first a system that has a Hamiltonian operator $\hat{H}$ with a discrete spectrum $\{ E_n\}$, and corresponding normalised eigenfunctions $\{\psi_n(x)\}$ satisfying 
\begin{equation} \label{eig}
\hat{H}\psi_n = E_n\psi_n ,~~~~~~\int dx\,\psi^*_m(x)\psi_n(x) = \delta_{mn} . \end{equation}
A general state of the system at any time $t$ is then specified by
\begin{equation} \label{state} 
\psi(x,t) = \sum_n c_n e^{-iE_nt/\hbar} \psi_n(x) = e^{-i\hat{H}t/\hbar}\psi(x,0) , 
\end{equation}
where the coefficients $c_n$ are any set of complex numbers satisfying the normalisation condition
\begin{equation} \label{norm}
\sum_n |c_n|^2 = 1 . 
\end{equation}

It follows immediately from Eqs.~(\ref{eig}) and (\ref{state}) that one has the identity
\begin{equation} \label{se}
\left[ \hat{H} - i\hbar (\partial/\partial t) \right] \psi(x,t) = 0 
\end{equation}
for all states of the system (in particular, one may apply the operator in square brackets to each term of the summation to obtain the result).  This equation is, of course, the Schr\"{o}dinger equation for the system.  However, one {\it cannot} in general rewrite Eq.~(\ref{se}) in the more familiar form
\begin{equation} \label{sefam}  
i\hbar (\partial/\partial t)\psi(x,t) = \hat{H}\psi(x,t).
\end{equation}
It is this somewhat subtle point, the inequivalence of Eqs.~(\ref{se}) and (\ref{sefam}) for certain states, that underlies the main results of this paper.

As a simple example, consider the case of a particle of mass $m$ confined to a one-dimensional box.  If the particle is confined to the interval $[0,L]$, with $\hat{H}=-\hbar^2/(2m)(d/dx)^2$ and the usual (Dirichlet) boundary conditions $\psi(x,t)=0$ at $x=0$ and $x=L$, then the energy eigenfunctions and eigenvalues are well known to be given by
\begin{equation} \label{sin} 
\psi_n(x) = (2/L)^{1/2}\sin n\pi x/L ,~~~E_n=(n\pi\hbar)^2/(2mL^2),~~~n=1,2,3,\dots . 
\end{equation}
For the particular case where the wavefunction is initially uniform over the interior of the box, one then has
\begin{equation} \label{box} 
c_n = \int_0^L dx\, \psi^*_n(x)\,\psi(x,0) = \frac{2\sqrt{2}}{\pi n},~~~~~~n=1,3,5,\dots ,
\end{equation}
with $c_n=0$ for $n=2,4,6,\dots$.  Hence, at time $t=0$, 
\[ \hat{H}\psi(x,0) = \sum_n c_nE_n\psi_n(x) = 
\frac{2\pi\hbar^2}{mL^{5/2}} \sum_{k=0}^\infty (2k+1)\sin \frac{(2k+1)\pi x}{L} , \]
which diverges for all $x\in(0,L)$.  Thus, 
Eq.~(\ref{sefam}) is meaningless for this example: the operator 
$\hat{H}$ acts not only to kick the wavefunction out of the Hilbert space, but to knock it right out of the set of functions altogether.

An analogous example of the inequivalence of Eqs.~(\ref{se}) and (\ref{sefam}), for the case of a {\it continuous} energy spectrum, is provided by a one-dimensional free particle of mass $m$ initially confined to some interval, i.e., with Hamiltonian operator $\hat{H} = -\hbar^2(d/dx)^2/(2m)$ and initial wavefunction
\begin{equation} \label{slit}
\psi(x,0) = L^{-1/2}e^{ip_0x/\hbar},~~~~~~-L/2<x<L/2 
\end{equation}
(corresponding, for example, to a plane wave incident on a one-dimensional slit).  It follows that the wavefunction at any later time has the Fourier decomposition
\[ \psi(x,t) = \left(\frac{2\hbar}{\pi L}\right)^{1/2} \int dp\, 
\frac{ \sin (p-p_0)L/2\hbar }{p-p_0} e^{ipx/\hbar-ip^2t/(2\hbar m)} , \]
and hence that $\hat{H}\psi(x,t)$ is not well-defined (in particular, the Fourier integrand of this quantity scales as $|p|$ for large $|p|$).

Thus, for some states, Eqs.~(\ref{se}) and (\ref{sefam}) are not equivalent - indeed, the latter equation has no meaning for these states.  It is this fact that lies behind the incompleteness of trajectory-based interpretations, as will be seen in section 4.  First, however, this subtlety of the Schr\"{o}dinger equation will be investigated a little further, in the following section.

Finally, it is of interest to note that the above examples of inequivalence arise with respect to the {\it position} representation of the quantum state, which is of course the representation having fundamental physical significance in trajectory-based interpretations.  In contrast, no analogous inequivalence arises for the Schr\"{o}dinger equation in the energy and the momentum representations, for either of the examples given above.  Moreover, it may be noted from 
Eq.~(\ref{state}) that, even in the position representation, the action of the unitary evolution operator $\hat{U}(t)=e^{-i\hat{H}t/\hbar}$ is always well defined, even though the action of the Hamiltonian $\hat{H}$ is not.  Thus, as per Wigner's theorem \cite{wigner} (and forming a basic element in most axiomatic approaches to quantum mechanics), it is unitary evolution which is fundamental to describing evolution on Hilbert space, with the Schr\"{o}dinger equation following as a secondary consequence.

\section{Energy considerations}

It is not difficult to see that the average energy $\langle H\rangle$ of the above two examples is infinite.  Thus, while these examples are perfectly valid quantum states, it is difficult to conceive of any method for their physical preparation.  Any interpretation that fails to explain them therefore suffers from a formal rather than a physical incompleteness.  Hence it is important to explore the issue of energy requirements further, and in particular to determine whether examples having {\it finite} average energies exist.

It is convenient for this purpose to return to the case of a discrete energy spectrum (similar considerations apply to the continuous case), and suppose that the amplitudes, energy eigenvalues, and eigenfunctions  scale respectively as
\begin{equation} \label{scale} 
|c_n| \sim n^{-\alpha},~~~~~E_n \sim n^\beta,~~~~~|\psi_n|\sim n^\gamma
\end{equation}
for large $n$, with $\alpha >1/2$ (to ensure that the state is square-integrable).  For the examples in section 2 one has $\alpha=1$, $\beta=2$ and $\gamma=0$.

It follows that (ignoring unimportant
 phase factors in the first line)
\begin{eqnarray*}
\hat{H}\psi &=& \sum_n c_nE_n\psi_n \sim \sum_n n^{\beta-\alpha +\gamma}, \\
\langle H\rangle &=& \sum_n |c_n|^2 E_n \sim  \sum_n n^{\beta-2\alpha}.
\end{eqnarray*}
Hence, one can arrange for $\hat{H}\psi$ diverge almost everywhere, while keeping the average energy $\langle H\rangle$ finite, by choosing $\beta-\alpha+\gamma>0$ and $\beta-2\alpha<-1$ respectively.  This is equivalent to the condition
\begin{equation} \label{cond}
(1+\beta)/2 < \alpha < \beta+\gamma
\end{equation}
on $\alpha$ (where for consistency one requires that $\beta>1-2\gamma$).  Eqs.~(\ref{scale}) and (\ref{cond}) provide a large parameter range corresponding to examples of particles with finite average energies for which Eq.~(\ref{sefam}) is generally invalid.  

Note that the form of the Schr\"{o}dinger equation in 
Eq.~(\ref{sefam}) must necessarily be valid whenever $\hat{H}\psi$ happens to be a member of the Hilbert space in question (and hence is well defined), i.e., whenever 
\[
\langle H^2\rangle = \int dx\, |\hat{H}\psi|^2 < \infty .
\]
Thus all counterexamples must have an infinite expectation value for the {\it square} of the energy of the system.  It follows immediately from Eqs.~(\ref{scale}) that $\beta-\alpha<-1/2$, leading via 
Eq.~(\ref{cond}) to the condition $\gamma<1/2$
for the asymptotic scaling of discrete energy eigenfunctions for any counterexample.

Finally, to give an optical example, consider a single-mode field of frequency $\omega$ in a nonlinear Kerr medium, with photon annihilation operator $\hat{a}$, number operator $\hat{N}=\hat{a}^\dagger\hat{a}$, and Hamiltonian operator
\[ \hat{H} = \hbar\omega\hat{N} + \kappa\hat{N}^2 . \]
The energy eigenfunctions $\psi_n(x)$, in the usual quadrature representation defined by $\hat{X}=(\hat{a}+ \hat{a}^\dagger)/2$, are Hermite-Gaussians, and for large photon numbers scale as $n^{-1/4}$ on any finite interval \cite{fractal}.  Further, for this case one has $E_n\sim n^2$.  Thus $\beta=2$ and $\gamma=-1/4$.  It then follows from 
Eq.~(\ref{cond}) that amplitudes scaling as 
\begin{equation} \label{opt}
|c_n|\sim n^{-\alpha},~~~~~~~3/2<\alpha<7/4, 
\end{equation}
yield states with finite average energy, for which the form of the 
Schr\"{o}dinger equation in Eq.~(\ref{sefam}) is not valid. As a particular example, one may choose the initial state of the field to have the number state expansion
\begin{equation} \label{optic}
|\psi_0\rangle := [\zeta(13/4)]^{-1/2} \sum_{n=0}^\infty (n+1)^{-13/8} |n\rangle ,
\end{equation}
corresponding to $\alpha=13/8$, where $\zeta(z)$ denotes the Riemann zeta-function and $|n\rangle$ denotes the $n$-th photon number eigenstate.  It would be of interest to find a scheme for the physical generation of such states.

\section{Incompleteness}

Given any wavefunction $\psi(x,t)$ associated with some quantum system, one can define quantities $P(x,t)$ and $S(x,t)$ via the polar decomposition 
\begin{equation} \label{polar}
\psi=P^{1/2}e^{iS/\hbar}
\end{equation} 
of the wavefunction.  Note that $P(x,t)=|\psi(x,t)|^2$ is the probability density associated with finding the system at position $x$ in configuration space, at time $t$.  

In trajectory-based interpretations of quantum mechanics, the position of the system is assumed to be a `real' property at all times, and the probability density $P(x,t)$ reflects incomplete knowledge of this property (in the de~Broglie-Bohm interpretation $P(x,t)$ also has a more fundamental role as a `real' physical degree of freedom associated with the guiding wave $\psi(x,t)$).  Thus quantum mechanics is interpreted as describing an {\it ensemble} of systems, with each member of the ensemble following a specific trajectory in configuration space.  Such a combination of trajectories and statistics provides an illuminating quasi-classical picture of quantum systems (although it should be noted that the trajectories can have rather `surrealistic' properties, that conflict with naive classical notions of position and its measurement \cite{scully}).  The aim of such interpretations is to explain the evolution of the ensemble as a consequence of the evolution of $P$ and $S$.  It is necessary to do this for {\it all possible} wavefunctions if such interpretations are to provide a {\it complete} explanation of quantum systems.

Now, if one assumes that the form of the Schr\"{o}dinger equation in Eq.~(\ref{sefam}) is valid, then for the Hamiltonian operator $\hat{H}=-(\hbar^2/2m)\nabla^2 +V(x)$ one may multiply this equation on the left by $\psi^*$, and take real and imaginary parts, to obtain the corresponding equations of motion
\begin{equation} \label{bohm}
\frac{\partial P}{\partial t} + \nabla .\left( P\,\frac{\nabla S}{m} \right) =0,~~~~~~
\frac{\partial S}{\partial t} + \frac{|\nabla S|^2}{2m} + V - 
\frac{\hbar^2 \nabla^2 P^{1/2}}{2mP^{1/2}} =0 ,
\end{equation}
for $P$ and $S$.  The first equation is a continuity equation, ensuring conservation of probability, and the second equation is a modified Hamilton-Jacobi equation.  Both the de~Broglie-Bohm interpretation and Nelson's stochastic interpretation are based on these equations
\cite{bohm,holland,nelsonpaper,nelsonbook}, and hence the consistency of these interpretations with quantum mechanics follows whenever Eq.~(\ref{sefam}) is valid.  

However, as was demonstrated by explicit example in the previous sections, there are perfectly well defined quantum states for which 
Eq.~(\ref{sefam}) is {\it not} valid.  For these states one {\it cannot} follow the above procedure to derive corresponding continuity and modified Hamilton-Jacobi equations. It follows that interpretations relying on these equations cannot explain the evolution of such states, and so are incomplete.  This result in fact applies not only to trajectory-based interpretations, but to {\it any} interpretation based on Eqs.~(\ref{bohm}), including the hydrodynamic and exact uncertainty interpretations \cite{holland,hydro,exact}.

Note that there seems at first to be a possible caveat on the above incompleteness result.  One could argue in particular that such interpretations don't {\it need} a corresponding Schr\"{o}dinger equation  - they only need Eqs.~(\ref{bohm}).  However, such an argument is consistent if and only if Eqs.~(\ref{bohm}) lead to the same predictions as the standard quantum formalism, for the states in question.  Unfortunately, this is not the case.  In particular, if the continuity and modified Hamilton-Jacobi equations are assumed to be {\it a priori} valid for such states, then the spatial and temporal derivatives of $P^{1/2}$ and $e^{iS/\hbar}$ must exist almost everywhere, and one can then {\it derive} Eq.~(\ref{sefam}) from 
Eqs.~(\ref{polar}) and (\ref{bohm}).  This contradicts the examples of the previous sections. It is therefore concluded that the continuity and modified Hamilton-Jacobi equations do {\it not} correctly describe the states in question.  

For the first example discussed in section 2, of a particle confined to a one-dimensional box with the wavefunction initially uniform over the interior of the box, the incompleteness of trajectory-based interpretations may be seen even more directly.  In particular, in such interpretations the initial wavefunction corresponds to an ensemble of particles with initial positions uniformly spread over the interior of the box (with no particles located at the boundaries of the box).  Thus, in the neighbourhood of every member of the ensemble, $P$ and $S$ are initally constant, implying that their spatial derivatives vanish.  It then follows immediately from Eqs.~(\ref{bohm}) that $P$ and $S$ must remain constant everywhere in the interior of the box - i.e., that the ensemble is {\it stationary}.  This contradicts the quantum evolution, where the wavefunction $\psi$ evolves as per Eq.~(\ref{state}).

The above example further provides an instance of the breakdown of the velocity equation
\begin{equation} \label{vel} 
v = m^{-1}\nabla S = (\hbar/m)\,\, {\rm Im}\,\left\{ \psi^{-1}\nabla\psi \right\},  
\end{equation}
postulated for each trajectory in the de~Broglie-Bohm interpretation \cite{bohm,holland}, and for the average drift velocity of the forward and backward processes in Nelson's stochastic interpretation 
\cite{nelsonpaper, nelsonbook}.  In particular, it has been shown by Berry that the spacetime dependence of the wavefunction $\psi(x,t)$ for this example has a {\it fractal} structure \cite{berry}.  Hence $\nabla \psi$ is not defined for almost all $x$ and $t$, and therefore Eq.~(\ref{vel}) cannot be used to define any corresponding trajectories or processes.

\section{Fractal connections}

The formal cause of the incompleteness of trajectory-based (and other) interpretations is seen to stem from the fact that the form of the 
Schr\"{o}dinger equation in Eq.~(\ref{sefam}) is not valid for all states - each side of this equation can be strongly divergent in the position representation, even for states with finite average energy.  Here some evidence is collected suggesting that, for quantum particles, this divergence is associated with fractal structures of the corresponding wavefunctions.

For the example of the one-dimensional particle in a box, with a wavefunction initially uniform over the box, Berry has shown that the probability distribution $P(x,t)$ has fractal dimension $3/2$ in the spatial direction for almost all fixed times $t$, and fractal dimension $7/4$ in the time direction for almost all fixed positions $x$. A simplified expression for $P(x,t)$ for this example is given in 
Ref.~\cite{carpet}, and an approximate experimental realisation of the fractal structure, via an optical analogue, is discussed in 
Refs.~\cite{opt1,opt2}.  Further fractal and near-fractal wavefunctions have been constructed by Wojcik et al.~\cite{fractal} and by Amanatidis et al.~\cite{lattice} (all having infinite average energies in the fractal limit).  For such wavefunctions both $P$ and $S$ in 
Eq.~(\ref{polar}) are typically also fractals, and hence provide further examples where Eqs.~(\ref{bohm}) and (\ref{vel}) are not well defined .

The fractal nature of the wavefunctions in Refs.~\cite{berry,fractal} was derived as a consequence of the result that functions of the form
\[ f(x) = \sum_n a_n e^{inx} , \]
for which the amplitudes $a_n$ scale asympotically as
\[ |a_n| \sim |n|^{-z}~~~~~~~~{\rm with}~1/2<z\leq 3/2, \]
are continuous but nondifferentiable, and have fractal dimension $5/2-z$ \cite{berry}.  Thus, for example, for a particle in a one-dimensional box with amplitudes $|c_n|\sim n^{-\alpha}$ in Eq.~(\ref{state}), it follows via 
Eq.~(\ref{sin}) that the wavefunction has a fractal structure whenever $1/2<\alpha\leq 3/2$  (this includes the particular case of the initially uniform wavefunction, for which $\alpha=1$).  Note, however, that since $\langle H\rangle\sim \sum_n n^{2-2\alpha}$, all such examples have infinite average energies and hence cannot be physically prepared.

It turns out that fractal structures can also be associated with states having finite average energies.  In particular, consider again states of the particle in the box with coefficients $|c_n|\sim n^{-\alpha}$ in 
Eq.~(\ref{state}), where $\alpha$ is now chosen to be in the parameter range corresponding to Eq.~(\ref{cond}) in section 3.  Since $\beta=2$ and $\gamma=0$ for this case, this range is given by $3/2<\alpha<2$.  Now, by construction, the corresponding states have finite average energy and do not satisfy the form of the Schr\"{o}dinger equation in Eq.~(\ref{sefam}).  On the other hand, they do not satisfy the fractal criterion given above.  However, from Eq.~(\ref{sin}) one finds that the {\it spatial derivative} of the corresponding wavefunctions has the form
\[ (d/dx)\psi(x,t) = \sqrt{2}\,\pi L^{-3/2} \sum_n nc_n \cos n\pi x/L .\]
This quantity {\it does} satisfy the above fractal criterion whenever $1/2<\alpha-1\leq 3/2$, and hence in particular for the range $3/2<\alpha<2$ of interest.   The corresponding fractal dimension is $7/2-\alpha$, and hence also lies between 3/2 and 2.  

Based on the above results, it is conjectured that the incompleteness of trajectory-based interpretations for quantum particles corresponds to the existence of states for which the spacetime dependence of the wavefunction, or of its spatial derivative, has a fractal structure.  

\section{Conclusions}

The incompleteness of trajectory-based interpretations arises for systems with unbounded Hamiltonian operators.  It applies not only to the very simple case of a wavefunction initially uniform over the interior of a box, but also to a number of examples having finite average energies and hence which can, in principle, be physically prepared.  For quantum particles the incompleteness of such interpretations appears to be connected with associated fractal structures.

Strictly speaking, one should differentiate here between the notions of formal and physical incompleteness.  The results of the paper show that trajectory-based interpretations are {\it formally} incomplete, as they do not describe all possible states in the Hilbert space, even though these states and their (unitary) evolution are well defined.  The results suggest that such interpretations are also {\it physically} incomplete, as some of the states in question have finite average energies, and so can plausibly be physically prepared from finite resources.  However, it is open to proponents of such interpretations to argue for physical completeness on the grounds that even the finite-average-energy counterexamples are unphysical.  For example, noting the discussion in section 3, it would suffice to provide a convincing argument that {\it all} moments of the energy of a physical system must be finite.  

In the case of wavefunctions for which $P$ and $S$ are fractals there would appear to be little chance of overcoming incompleteness, via some supplementary rule that specifies how to generate the corresponding (presumably fractal) trajectories.  In contrast, in the case of wavefunctions for which the spatial derivative is a fractal, the velocity equation in Eq.~(\ref{vel}) is well defined, and hence might be used to specify a set of associated trajectories (eg, via $\dot{x}=v$ in the de~Broglie-Bohm interpretation).  However, given the nonexistence of a corresponding continuity equation as per Eq.~(\ref{bohm}), it is not clear that an ensemble of these trajectories can evolve in agreement with the 
Schr\"{o}dinger equation in Eq.~({\ref{se}).  It would be of interest to perform some numerical experiments in this regard.

As previously remarked, the incompleteness result applies to any interpretation that relies on the continuity and modified Hamilton-Jacobi equations in Eq.~(\ref{bohm}) \cite{bohm,holland,nelsonpaper, nelsonbook,hydro,exact}.  Thus quantum mechanics goes where these intepretations do not follow, despite their (at least in principle) duty to do so.

\end{document}